\documentclass[conference,10pt]{IEEEtran}
\usepackage{graphicx}
\usepackage{amsmath}
\usepackage{mathtools}
\usepackage{amssymb}
\usepackage{math_symbols}
\usepackage{calc}
\usepackage{booktabs,hhline} 
\usepackage{enumitem}
\usepackage{siunitx}
\usepackage{acronym}
\usepackage[english]{babel}
\usepackage{times}
\usepackage{url}
\usepackage{algorithm}
\usepackage{tikz}
\usetikzlibrary{calc,arrows,positioning}
\usepackage{tabularx}
\acrodef{ssm}[SSM]{state-space model}
\acrodef{ss}[SS]{state-space}
\acrodef{kf}[KF]{Kalman filter} 
\acrodef{pf}[PF]{particle filter} 
\acrodef{kg}[KG]{Kalman gain}
\acrodef{ekf}[EKF]{extended \ac{kf}}
\acrodef{ckf}[CKF]{cubature \ac{kf}}
\acrodef{ukf}[UKF]{unscented \ac{kf}}
\acrodef{brr}[BRR]{Bayesian recursive relation} 
\acrodef{pdf}[PDF]{probability density function}
\acrodef{cke}[CKE]{Chapman-Kolmogorov equation}
\acrodef{gmf}[GMF]{Gaussian mixture filter}
\acrodef{gm}[GM]{Gaussian mixture}
\acrodef{gf}[GF]{Gaussian filter}
\acrodef{gpb}[GPB]{generalized pseudo-Bayesian}
\acrodef{imm}[IMM]{interacting multiple-model}
\acrodef{ut}[UT]{unscented transform}
\acrodef{mc}[MC]{Monte Carlo}
\acrodef{rmse}[RMSE]{root mean-squared error}
\acrodef{mse}[MSE]{mean-squared error}
\acrodef{pmf}[PMF]{point-mass filter}
\acrodef{kld}[KLD]{Kullback-Leibler divergence}
\acrodef{sde}[SDE]{Shanon differential entropy}
\acrodef{ai}[AI]{artificial intelligence} 
\acrodef{dnn}[DNN]{deep neural network} 
\acrodef{cnn}[CNN]{convolutional neural network} 
\acrodef{nn}[NN]{neural network}  
\acrodef{gnn}[GNN]{graph neural network} 
\acrodef{rnn}[RNN]{recurrent neural network} 
\acrodef{fc}[FC]{fully connected} 
\acrodef{snr}[SNR]{signal-to-noise ratio}
\acrodef{ml}[ML]{machine learning}
\acrodef{gru}[GRU]{gated recurrent unit} 
\acrodef{lstm}[LSTM]{long short-term memory} 
\acrodef{rkn}[RKN]{recurrent Kalman network}
\acrodef{pbm}[PBM]{physics-based model}
\acrodef{apbm}[APBM]{data-augmented physics-based model}
\acrodef{tm}[TM]{true model}
\acrodef{rts}[RTS]{Rauch-Tung-Striebel}
\acrodef{ddm}[DDM]{data-driven model}
\acrodef{vae}[VAE]{variational autoencoder}
\acrodef{pinn}[PINN]{physics-informed neural network}

\acrodef{narma}[NARMA]{nonlinear autoregressive moving-average}
\acrodef{armax}[ARMAX]{autoregressive moving-average model}
\acrodef{ungm}[UNGM]{univariate nonstationary Gaussian model}

\acrodef{nat}[NaT]{navigation and tracking}

\IEEEoverridecommandlockouts

\newcommand\clearrow{\global\let\rowmac\relax}
\clearrow

\usepackage{cite} 

\def\fsg{\mathrm{fsg}}

\def\rmse{\mathrm{RMSE}}
\def\pf{\mathrm{pf}}

\acrodef{psg}[PSG]{predicted state grid}
\acrodef{fsg}[FSG]{filtered state grid}

\begin{document}

\title{Efficient Gaussian Mixture Filters based on Transition Density Approximation}

\author{%
\IEEEauthorblockN{Ond\v{r}ej Straka}
\IEEEauthorblockA{%
European Centre of Excellence NTIS, \\ 
  University of West Bohemia in Pilsen, Czech Republic\\ E-mail: straka30@kky.zcu.cz}
  \and
\IEEEauthorblockN{Uwe D. Hanebeck} 
  \IEEEauthorblockN{
  Intelligent Sensor-Actuator-Systems Laboratory (ISAS)\\
Karlsruhe Institute of Technology
Karlsruhe, Germany\\ E-mail: uwe.hanebeck@kit.edu}
\thanks{ The work was partially supported by the Ministry of Education, Youth and Sports of the Czech Republic under project ROBOPROX - Robotics and Advanced Industrial Production CZ.02.01.01/00/22\_008/0004590.}
}%
\selectlanguage{english}
\maketitle
\begin{abstract}
Gaussian mixture filters for nonlinear systems usually rely on severe approximations when calculating mixtures in the prediction and filtering step. Thus, offline approximations of noise densities by Gaussian mixture densities to reduce the approximation error have been proposed. This results in exponential growth in the number of components, requiring ongoing component reduction, which is computationally complex. In this paper, the key idea is to approximate the true transition density by an axis-aligned Gaussian mixture, where two different approaches are derived. These approximations automatically ensure a constant number of components in the posterior densities without the need for explicit reduction. In addition, they allow a trade-off between estimation quality and computational complexity.
\end{abstract}

\section{Introduction}\label{sec:introduction}
State estimation is a key element in many areas, such as tracking, guidance, positioning, navigation, sensor fusion, fault detection, and decision-making.
Its goal is to estimate the state of a dynamic system, which is, in general, not directly measurable, from a set of noisy measurements.

State estimation began in the sixties with the introduction of the \ac{kf}, which optimizes state estimation for linear systems by minimizing the mean square error.
Its structure is followed by many algorithms addressing state estimation of nonlinear systems.
The algorithms following the optimization approach provide point state estimates and covariance matrices of estimate errors.
Another approach to state estimation uses \acp{brr}, incorporating a Bayes equation and \ac{cke}, to provide the state estimate as a conditional \ac{pdf} based on a set of measurements.
The conditional \ac{pdf} completely describes the estimated state; however, the \acp{brr} are analytically tractable only for a few special cases.

There are many approximate solutions to the \acp{brr} leading to algorithms of various complexity.
Assuming joint \ac{pdf} of the state and measurement prediction Gaussian leads to a group called \acp{gf} (e.g., cubature filter~\cite{ArHa:09} or the stochastic integration filter~\cite{DuStrSi:13}), which are computationally light.
However, the assumption seldom holds, and thus, \ac{gf} performance for strongly nonlinear systems is poor.

Another group of approximate Bayesian algorithms results from discrete approximation of the posterior \ac{pdf}.
These include the \acp{pmf}~\cite{SiKraSo:06} replacing the continuous support of the posterior by a grid of weighted points, which is usually orthogonal and equidistant.
Such a grid is flexible in representing the posterior but results in computationally demanding algorithms due to the convolution calculated in the \ac{cke}.
\Acp{pf}~\cite{RiArGo:03} belong to the same group, but they replace the continuous support with a set of randomly positioned samples.
They are computationally lighter than the \acp{pmf}, but the estimates are subject to random effects.

\Acp{gmf}~\cite{sorensonRecursiveBayesianEstimation1971}
are Bayesian algorithms filling the gap between \acp{gf} and \acp{pmf}.
The support of the posterior is continuous since they represent it by a mixture of Gaussian \acp{pdf}.
The usage of a \ac{gm} instead of a single Gaussian \ac{pdf} results in better accuracy, even for highly nonlinear problems. This improvement occurs because each \ac{gm} term has a small variance compared to the single Gaussian \ac{pdf}, which mitigates the impact of strong nonlinearity.
Compared to \acp{pf} and \acp{pmf}, the \acp{gmf} are usually computationally lighter.
For arbitrary nonlinear state and measurement equations, \acp{gmf} can be classified into two different approaches: local ones and global ones.

\emph{Local approaches} process the \ac{gm} components individually by a bank of filters, each processing a single component. This has been pioneered in \cite{sorensonRecursiveBayesianEstimation1971} with the first use of \acp{gm} for Bayesian nonlinear filtering. 
The unavoidable increase in the number of components due to noises described by \acp{gm} is handled by simple neglecting and combination.
In \cite{itoGaussianFiltersNonlinear2000}, 
a bank of Gaussian filters for the independent updates of Gaussian components is proposed with three different rules to update the component weights after the correction step.

A more recent discussion of weights calculation after the correction step based on prior linearization, posterior linearization, and without any linearization can be found in \cite{AAS25_Durant}.
More complex individual filters have been used in \cite{kotechaGaussianSumParticle2003}, 
where a component-wise update is performed using a bank of Gaussian particle filters as proposed in \cite{kotechaGaussianParticleFiltering2003}.

\emph{Global approaches} jointly process all the mixture components.
An early global approach is the update step for prior \ac{gm} with arbitrary nonlinear measurement equation in \cite{CDC03_Feiermann-ProgBayes}. Instead of solving the complex problem of approximating the prior \ac{gm}, a system of ordinary differential equations for posterior mixture parameters is solved over an artificial time interval.
A recent global approach~\cite{FUSION24_Frisch}
considers the update step for a Gaussian mixture particle filter with deterministic samples. It works by drawing unweighted samples from the prior \ac{gm}, assigning weights from the likelihood, computing higher-order moments from this sample-based posterior, and finally determining the posterior \ac{gm} from moments while minimizing its Fisher information.
This approach is based on calculating the Fisher information for \acp{gm} in \cite{ACC24_Hanebeck}.

Alternatively, global processing can be achieved with transition densities approximated by \acp{gm}, e.g., \cite{willsNumericallyRobustBayesian2023}.
In general, for arbitrary mixtures, this leads to an exponential increase in the number of components over time in both the filtering and prediction steps.
Usually, this is taken care of by regular mixture reduction, which is a non-convex and computationally complex optimization problem.
To automatically limit the number of components in the prediction step with a \ac{gm} transition density, \cite{MFI06_Huber} proposed to perform a decomposition into axis-aligned components. This not only allows a prespecified number of components without any reduction but also leads to closed-form expressions for the posterior mixture.


This paper follows the global processing and aims to propose \ac{gmf} algorithms that impose the structure of the predictive and posterior \ac{pdf} through a decomposition of the transition \ac{pdf} to axis-aligned \acp{gm}.
This allows the user to control accuracy and computational complexity by specifying the decomposition.

The paper is structured as follows: Section~\ref{sec:bayesian_estimation} introduces the state estimation problem, the \acp{brr}, and their solution by the \ac{gmf}.
Section~\ref{sec:decomposition} describes two decompositions of the transition \ac{pdf} leading to two proposed \ac{gmf} algorithms described in Section~\ref{sec:GMM_TPD}.
A numerical illustration of the developed algorithms is given in Section~\ref{sec:numerical_illustration}, and Section~\ref{sec:conclusion} provides concluding remarks.
\section{Bayesian state estimation and Gaussian mixture filter}\label{sec:bayesian_estimation}
Consider a discrete-time stochastic system described by a nonlinear state-space model
\begin{subequations}\label{eq:sseq}
\begin{align}\label{eq:sseqx}
    \bfx_{k+1} &= \bff_k(\bfx_k) + \bfw_k,\\
    \bfz_k &= \bfh_k(\bfx_k) + \bfv_k,\label{eq:sseqz}
\end{align}
\end{subequations}
where $\bfx_k\in\real^{n_x}$ and $\bfz_k\in\real^{n_z}$ represent the immeasurable state of the system and the available measurement at time instant $k=0,1,\ldots$, respectively. The functions $\bff_k:\real^{n_x}\mapsto\real^{n_x}$ and $\bfh_k:\real^{n_x}\mapsto\real^{n_z}$ are assumed known. The state noise $\bfw_k\in\real^{n_x}$ and measurement noise $\bfv_k\in\real^{n_z}$ are described by known \acp{pdf} $p_{\bfw_k}$ and $p_{\bfv_k}$. The initial state $\bfx_0$ is given by known \ac{pdf} $p_{\bfx_0}$.
Both noises are assumed to be white, mutually independent, and independent of the initial state.

The model~\eqref{eq:sseq} can be expressed using transition \ac{pdf}~\eqref{eq:sspdfx} and measurement \ac{pdf}~\eqref{eq:sspdfz}
\begin{subequations}\label{eq:sspdf}
    \begin{align}\label{eq:sspdfx}
        p(\bfx_{k+1}|\bfx_k)&=p_{\bfw_k}(\bfx_{k+1}-\bff_k(\bfx_k))\\
        p(\bfz_{k}|\bfx_k)&=p_{\bfv_k}(\bfz_{k}-\bfh_k(\bfx_k)).\label{eq:sspdfz}        
    \end{align}
\end{subequations}
\subsection{Bayesian state estimation}
The goal of state estimation is to infer the posterior \ac{pdf} $p(\bfx_z|\bfz^k)$ of the state $\bfx_k$ given all the measurements available up to time~$k$ denoted as $\bfz^k\coloneqq[\bfz_1\T,\bfz_2\T,\cdots,\bfz_k\T]\T$.
The general solution to the problem is provided by the \acp{brr} consisting of the Bayes equation~\eqref{eq:brrbayes} and the \ac{cke}~\eqref{eq:brrcke}
\begin{subequations}\label{eq:brr}
    \begin{align}\label{eq:brrbayes}
        p(\bfx_k|\bfz^k)&=\frac{p(\bfz_k|\bfx_k)p(\bfx_k|\bfz^{k-1})}{p(\bfz_k|\bfz^{k-1})}\\
        p(\bfx_k|\bfz^{k-1})&=\int p(\bfx_k|\bfx_{k-1})p(\bfx_{k-1}|\bfz^{k-1})\d\bfx_{k-1},\label{eq:brrcke}
    \end{align}
\end{subequations}
where $p(\bfz_k|\bfz^{k-1})$ is the evidence given by $p(\bfz_k|\bfz^{k-1})=\int p(\bfz_k|\bfx_k)p(\bfx_k|\bfz^{k-1})\d\bfx_k$.
The initial condition for the \acp{brr} is $p(\bfx_0|\bfz^0)=p(\bfx_0)$.
The calculation of the \acp{brr} thus involves alternating the filtering step~\eqref{eq:brrbayes} and the prediction step~\eqref{eq:brrcke}.
Usually, an approximate solution has to be used to obtain the filtering \ac{pdf} $p(\bfx_k|\bfz^k)$ and the predictive \ac{pdf} $p(\bfx_k|\bfz^{k-1})$.
\subsection{Gaussian mixture filter}
The \ac{gmf} assumes the predictive \ac{pdf} in the form of a \ac{gm}
\begin{align}\label{eq:gmf_initial}
    p(\bfx_{k}|\bfz^{k-1})=\sum_{i=1}^{N_{k|k-1}}\alpha_{k|k-1}^i\calN\{\bfx_k;\bfm^{\bfx,i}_{k|k-1},\bfSigma^{\bfx,i}_{k|k-1}\},
\end{align}
where the notation $\calN\{\bfx;\bfm,\bfSigma\}$ stands for the Gaussian \ac{pdf} of the random variable $\bfx$ with mean $\bfm$ and covariance matrix $\bfSigma$ and $\alpha_{k|k-1}^i\geq 0,\ i=1\ldots N_{k|k-1}$ are the weights that sum to one, i.e., $\sum_{i=1}^{N_{k|k-1}}\alpha_{k|k-1}^i=1$.
For \ac{gm}~\eqref{eq:gmf_initial}, Eq.~\eqref{eq:brrbayes} can be arranged as
\begin{align}\label{eq:gmfbayes}
        p(\bfx_k|\bfz^k)&=\frac{p(\bfz_k|\bfx_k)\sum_{i=1}^{N_{k|k}}\alpha_{k|k-1}^i\calN\{\bfx_k;\bfm^{\bfx,i}_{k|k-1},\bfSigma^{\bfx,i}_{k|k-1}\}}{p(\bfz_k|\bfz^{k-1})}\nonumber\\
        &=\sum_{i=1}^{N_{k|k-1}}\underbrace{\frac{\alpha_{k|k-1}^i p^i(\bfz_k|\bfz^{k-1})}{\sum_{j=1}^{N_{k|k-1}}\alpha_{k|k-1}^j p^j(\bfz_k|\bfz^{k-1})}}_{\alpha_{k|k}^i}\,p^i(\bfx_k|\bfz^k)\nonumber\\
        &=\sum_{i=1}^{N_{k|k-1}}\alpha_{k|k}^i\, p^i(\bfx_k|\bfz^k),
\end{align}
where $p^i(\bfx_k|\bfz^k)$ corresponds to calculation of~\eqref{eq:brrbayes} for $p(\bfx_k|\bfz^{k-1})=\calN\{\bfx_k;\bfm^{\bfx,i}_{k|k-1},\bfSigma^{\bfx,i}_{k|k-1}\}$ and similarly, $p^i(\bfz_k|\bfz^{k-1})=\int p(\bfz_k|\bfx_k)\calN\{\bfx_k;\bfm^{\bfx,i}_{k|k-1},\bfSigma^{\bfx,i}_{k|k-1}\}\d\bfx_k$.
To preserve the \ac{gm} form of the predictive \ac{pdf}, the following approximation is calculated for each~$i$:
\begin{align}\label{eq:gmf_posteriorlocal}
    p^i(\bfx_k|\bfz^k)\approx\calN\{\bfx_k;\bfm^{\bfx,i}_{k|k},\bfSigma^{\bfx,i}_{k|k}\}
\end{align}
by means of moment matching with Taylor series expansion, \ac{ut}, or cubature or quadrature rules.
The filtering step thus involves $N_{k|k-1}$ parallel calculations of~\eqref{eq:brrbayes} for  $p(\bfx_k|\bfz^{k-1})=\calN\{\bfx_k;\bfm^{\bfx,i}_{k|k-1},\bfSigma^{\bfx,i}_{k|k-1}\}$ and evaluation of $p^i(\bfz_k|\bfz^{k-1})$, for which the measurement prediction $\hbfz_{k|k-1}^i=\int \bfh_k(\bfx_k) \calN\{\bfx_k;\bfm^{\bfx,i}_{k|k-1},\bfSigma^{\bfx,i}_{k|k-1}\}\d\bfx_k$ is typically used yielding $p^i(\bfz_k|\bfz^{k-1})=p_{\bfv_k}(\bfz_k-\hbfz_{k|k-1}^i)$.
Now, having the \ac{gm} representation of the posterior \ac{pdf} 
\begin{align}\label{eq:gmf_posterior}
    p(\bfx_{k}|\bfz^{k})=\sum_{i=1}^{N_{k|k}}\alpha_{k|k}^i\calN\{\bfx_k;\bfm^{\bfx,i}_{k|k},\bfSigma^{\bfx,i}_{k|k}\}
\end{align}
with $N_{k|k}=N_{k|k-1}$, the predictive \ac{pdf} for the next time step $p(\bfx_{k+1}|\bfz^k)$ is calculated by the \ac{cke}
\begin{align}
    p(\bfx_{k+1}|\bfz^k) &= \int p(\bfx_{k+1}|\bfx_k) \sum_{i=1}^{N_{k|k}}\alpha_{k|k}^i\calN\{\bfx_k;\bfm^{\bfx,i}_{k|k},\bfSigma^{\bfx,i}_{k|k}\}\d\bfx_k\nonumber\\
    &= \sum_{i=1}^{N_{k|k}}\alpha_{k|k}^i\underbrace{\int p(\bfx_{k+1}|\bfx_k) \calN\{\bfx_k;\bfm^{\bfx,i}_{k|k},\bfSigma^{\bfx,i}_{k|k}\}\d\bfx_k}_{p^i(\bfx_{k+1}|\bfz^k)}\nonumber\\
    &=\sum_{i=1}^{N_{k+1|k}}\alpha_{k|k}^i\, p^i(\bfx_{k+1}|\bfz^k),
\end{align}
where $N_{k+1|k}=N_{k|k}$.
Again, the approximation
\begin{align}\label{eq:gmf_predictivelocal}
    p^i(\bfx_{k+1}|\bfz^k)\approx\calN\{\bfx_{k+1};\bfm^{\bfx,i}_{k+1|k},\bfSigma^{\bfx,i}_{k+1|k}\}
\end{align}
facilitates preservation of the \ac{gm} form of the predictive \ac{pdf}
\begin{align}\label{eq:gmf_predictive}
    p(\bfx_{k+1}|\bfz^{k})=\sum_{i=1}^{N_{k+1|k}}\alpha_{k+1|k}^i\calN\{\bfx_{k+1};\bfm^{\bfx,i}_{k+1|k},\bfSigma^{\bfx,i}_{k+1|k}\}.
\end{align}

The equalities $N_{k+1|k}=N_{k|k}=N_{k|k-1}$ suggest that the number of terms of the \ac{gm} approximations of predictive and posterior \acp{pdf} is fixed to $N_{0|0}$.
Such a situation results from the representation/approximation of the initial \ac{pdf}
\begin{align}\label{eq:gmf_prior}
    p(\bfx_{0}|\bfz^{0})=\!/\!\approx\sum_{i=1}^{N_{0|0}}\alpha_{0|0}^i\calN\{\bfx_{0};\bfm^{\bfx,i}_{0|0},\bfSigma^{\bfx,i}_{0|0}\}.
\end{align}
If the transition \ac{pdf} $p(\bfx_{k+1}|\bfx_k)$ in~\eqref{eq:sspdfx} has the \ac{gm} form with $W$ terms (due to several state dynamics models or process noise \ac{pdf} $p_{\bfw_k}$ being a \ac{gm}), then $N_{k+1|k}=W\cdot N_{k|k}$. 
If the measurement \ac{pdf} $p(\bfz_k|\bfx_k)$ in~\eqref{eq:gmf_posterior} is of \ac{gm} form with $V$ terms (due to several measurement models or measurement noise \ac{pdf} $p(\bfv_k)$ being a \ac{gm}), then $N_{k|k}=V\cdot N_{k|k-1}$.
In both cases, the number of \ac{gm} terms in the posterior and predictive \acp{pdf} grows exponentially.
To achieve tractable \ac{gmf} algorithms, techniques for merging or pruning terms of the \acp{gm} such as generalized pseudo-Bayesian or interacting multiple-model techniques~\cite{BaLiKi:01}.


\section{Decomposition of the transition density}\label{sec:decomposition}
The two \acp{gmf} proposed in this paper result from \emph{global processing}, in particular from an approximate decomposition of the transition \ac{pdf} using axis-aligned \acp{gm}. It will be shown that this results in the \emph{user-defined structure} of the \ac{gm} in the predictive \ac{pdf}. 
The structure can be adapted in relation to the current working point.
Such an approach can be used even for cases when the model and the initial \ac{pdf} $p(\bfx_0|\bfz^0)$ do not have a \ac{gm} structure.
The proposed \acp{gmf} differ in the specification of the user-defined structure.
The first \ac{gmf} algorithm defines the decomposition structure through the state described by the posterior (filtering) \ac{pdf} $p(\bfx_k|\bfz^k)$, i.e., the structure is defined by a \ac{fsg}.
The second \ac{gmf} algorithm defines the decomposition structure through the state described by the predictive \ac{pdf}  $p(\bfx_{k+1}|\bfz^k)$, i.e., the structure is defined through a \ac{psg}. 

For convenience, the decompositions will be introduced for a zero mean Gaussian state noise, i.e., 
\begin{align}\label{eq:transition_gaussian}
    p(\bfx_{k+1}|\bfx_{k})&= p_{\bfw_{k}}(\bfx_{k+1}-\bff_{k}(\bfx_{k}))=\calN\{\bfx_{k+1};\bff_{k}(\bfx_k),\bfQ\}
\end{align}
However, they can be designed for any $p_{\bfw_k}$. 
\subsection{Decomposition with filtered state grid}
\begin{figure}
    \centering
    \includegraphics[width=\linewidth]{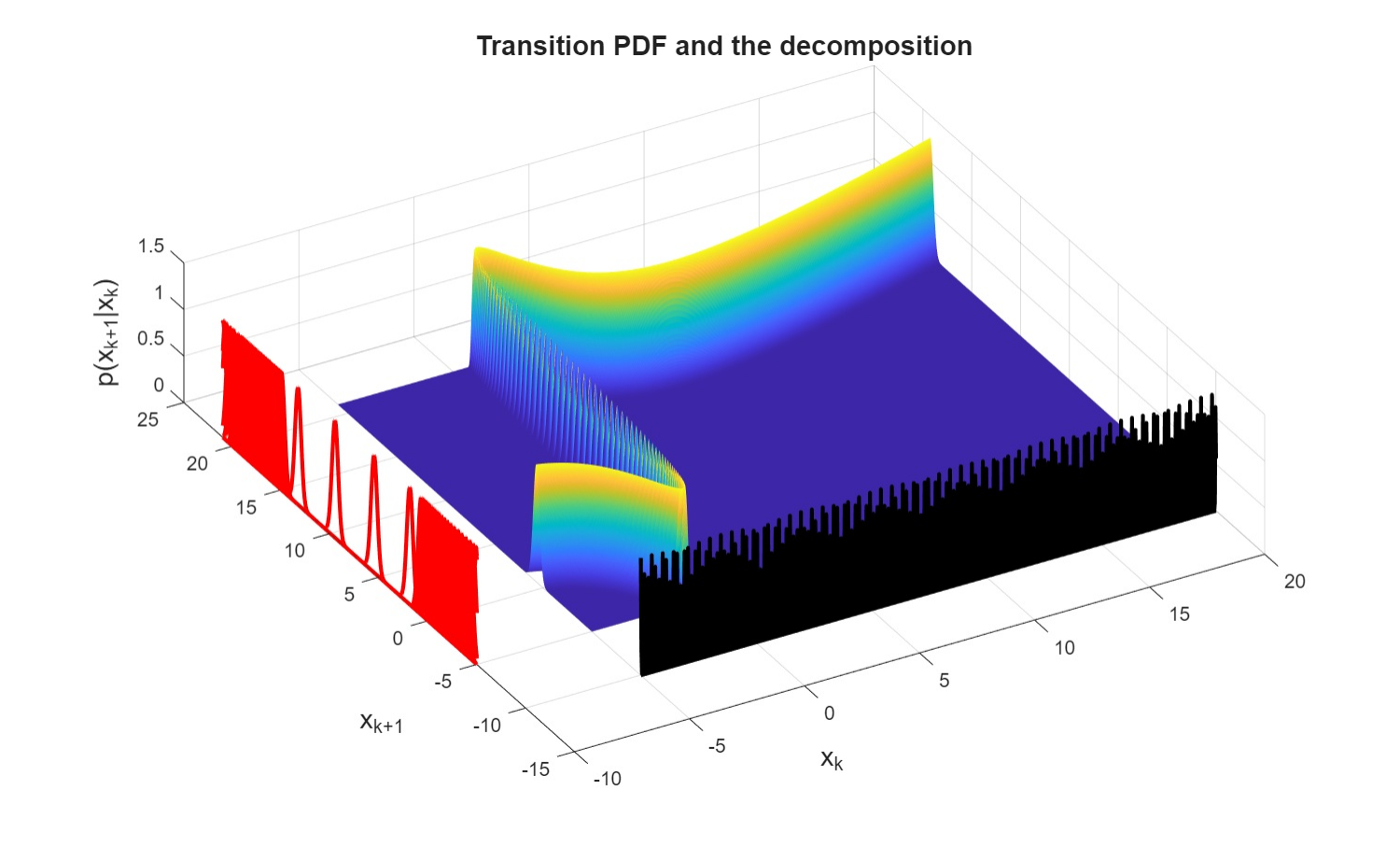}
    \caption{Illustration of the transition \ac{pdf} decomposition with \ac{fsg}. Red curves correspond to functions $\bfg$ while black curves to functions $\bfgamma$.}
    \label{fig:illustration_filtered}
\end{figure}

This decomposition described in~\cite{MFI06_Huber} assumes the transition density~\eqref{eq:transition_gaussian} to be decomposed as
\begin{multline}\label{eq:decomposition_filtered}
    p(\bfx_{k+1}|\bfx_{k})\approx \hp^\fsg(\bfx_{k+1}|\bfx_{k}) =\\ \sum_{j=1}^{M_{k+1}}\omega^j_k\, \bfg^j(\bfx_{k+1};\bftheta^{\bfg,j}_{k+1})\,\bfgamma^j(\bfx_{k};\bftheta^{\bfgamma,j}_k),
\end{multline}
where functions $\bfg^j$ and $\bfgamma^j$ are given by Gaussian \acp{pdf}
\begin{subequations}\label{eq:fsg_decomposition}
\begin{align}\label{eq:fsg_decomposition_xkpo}
    \bfg^j(\bfx_{k+1};\bftheta_{k+1}^{\bfg,j})&=\calN\{\bfx_{k+1};\bfm^{\bfg,j}_{k+1},\bfSigma^{\bfg,j}_{k+1}\}\\
 \label{eq:fsg_decomposition_xk}
    \bfgamma^j(\bfx_{k};\bftheta_{k}^{\bfgamma,j})&=\calN\{\bfx_{k};\bfm^{\bfgamma,j}_k;\bfSigma^{\bfgamma,j}_k\}
\end{align}
\end{subequations}
The decomposition defines the fixed structure of the predictive PDF $p(\bfx_{k+1}|\bfz^{k})$ through setting parameters $\bftheta^{\bfgamma,j}_{k},\, j=1\ldots , M_{k+1}$. 
The decomposition is illustrated in Figure~\ref{fig:illustration_filtered} for a scalar state transition density
\begin{align}\label{eq:ungm_transition}
    p(x_{k+1}|x_k)=\calN\{x_{k+1};0.5x_k+\tfrac{25x_k}{1+x_k^2}+8\cos(1.2k),Q\}
\end{align}
(for $k=0$) appearing in the highly nonlinear \ac{ungm} problem. 
\subsubsection{Calculation of the parameters} 
The parameters $\bftheta^{\bfg}_{k+1}\coloneqq\{\bftheta^{\bfg,j}_{k+1}\}_{j=1}^{M_{k+1}}$, $\bftheta^{\bfgamma}_{k}\coloneqq\{\bftheta^{\bfgamma,j}_{k}\}_{j=1}^{M_{k+1}}$ and the weights $\omega_k\coloneqq\{\omega^j_k\}_{j=1}^{M_{k+1}}$ are calculated to minimize the squared error between the transition \ac{pdf} and its approximation~\eqref{eq:decomposition_filtered}
\begin{multline}\label{eq:fsg_criterion}
    J(\bftheta^{\bfg}_{k+1},\bftheta^{\bfgamma}_{k},\omega_k)=\\\iint \left(p(\bfx_{k+1}|\bfx_k)- \hp^\fsg(\bfx_{k+1}|\bfx_k)\right)^2\d\bfx_{k+1}\d\bfx_k.
\end{multline}
The reason for choosing the integral of a squared difference between the transition \ac{pdf} and its approximation is the fact that for the Gaussian transition \ac{pdf}~\eqref{eq:transition_gaussian} and the approximation~\eqref{eq:decomposition_filtered}, some terms can be integrated analytically.
The large number of parameters makes the minimization of~\eqref{eq:fsg_criterion} intractable.
Hence, some parameters are prescribed by the user, and the optimization is performed only over a few.

First, the number of terms $M_{k+1}$ and the location parameters $\bfm^{\bfgamma,j}_{k}$ are selected to achieve good approximation of~\eqref{eq:transition_gaussian} in the non-negligible support of the posterior $p(\bfx_k|\bfz^k)$.
Second, the location parameters $\bfm^{\bfg,j}_{k+1}$ are calculated as
\begin{align}
    \bfm^{\bfg,j}_{k+1} = \bff_k(\bfm^{\bfgamma,j}_{k}),\, j=1,\ldots ,M_{k+1}.
\end{align}
Third, covariances $\bfSigma^{\bfg}_{k+1}$ are set as $\bfSigma^{\bfg,j}_{k+1}=\bfQ,\, j=1,\ldots,M_{k+1}$ due to~\eqref{eq:transition_gaussian}.
Fourth, to respect the shape of the function $\bff_k$ at the location $\bfm^{\bfgamma,j}_k$, the covariance $\bfSigma^{\bfgamma}_{k}$ is set to be proportional to
\begin{align}\label{eq:bfSigma}
    \bfSigma^{\bfgamma,j}_{k}\propto \bar{\bfSigma}^{\bfgamma}_{k}\left(\tfrac{\partial \bff_k(\bfx_k)}{\partial\bfx_k}\Big|_{\bfx_k=\bfm^{\bfgamma,j}_k}\!\right)^{-1}\!\bfQ\!\left(\tfrac{\partial \bff_k(\bfx_k)}{\partial\bfx_k}\Big|_{\bfx_k=\bfm^{\bfgamma,j}_k}\!\right)^{-\intercal},
\end{align}
where $(\bfF)^{-\intercal}$ stands for the transpose of the inverse of $\bfF$. The proportionality factor denoted as $\bar{\bfSigma}^{\bfgamma}_k$ is common for all $\bfSigma^{\bfgamma,j}_{k},\, j{=}1,\ldots,M_{k+1}$. Its value is obtained by minimizing~\eqref{eq:fsg_criterion}.
Finally, the weights $\omega_k$ are set as
\begin{align}\label{eq:omega}
    \omega^j_k\propto \bar{\omega}_k = 1/{\sqrt{(2\pi)^{n_x}|\bar{\bfSigma}^{\bfgamma,j}_{k}|}}
\end{align}
so that 
\begin{multline}
    p_{\bfx_{k+1}|\bfx_k}(\bfm^{\bfg,j}_{k+1}|\bfm^{\bfgamma,j}_k)=\\\bfomega^j_k \bfg^j(\bfm^{\bfg,j}_{k+1};\bftheta^{\bfg,j}_{k+1})\,\bfgamma^j(\bfm^{\bfgamma,j}_{k};\bftheta^{\bfgamma,j}_k) .
\end{multline}
The proportionality factor denoted as $\bar{\omega}_k$ is common for all $\omega^{j}_k,\, j=1,\ldots,M_{k+1}$. 
Its value is obtained by minimizing~\eqref{eq:fsg_criterion}.
Hence, the criterion is minimized w.r.t. the proportionality factors $\bar{\bfSigma}^{\bfgamma}_k$ and $\bar{\omega}_k$ both being scalars.
\subsubsection{Possible generalization}
When the function $\bff_k$ is time-varying, it may be necessary to recompute the decomposition for each time instant. In a special case, when the time instant acts in an additive way such as $\bff_{k+m}(\bfx)=\bff_k(\bfx_k)+\Delta_k$, where $\Delta_k:\real\mapsto\real^{n_x}$, the decomposition may be pre-computed only for a single time instant (e.g., $k=0$) and then shifted in direction of $\bfx_{k+1}$ by $\Delta_k-\Delta_0$.

The generalization to higher dimensions is straightforward. The number of location parameters $\bfm_k^{\bfgamma}$ may grow exponentially with the state dimension $n_x$. Optimization of the criterion~\eqref{eq:fsg_criterion} to obtain the covariance matrices $\bfSigma_k^{\bfgamma}$ and the weights $\omega$ may be difficult but an approach similar to~\eqref{eq:bfSigma} and~\eqref{eq:omega} may be used. The decomposition can also be designed for the non-Gaussian transition \ac{pdf}. 

\subsection{Decomposition with predicted state grid}
\begin{figure}
    \centering
    \includegraphics[width=\linewidth]{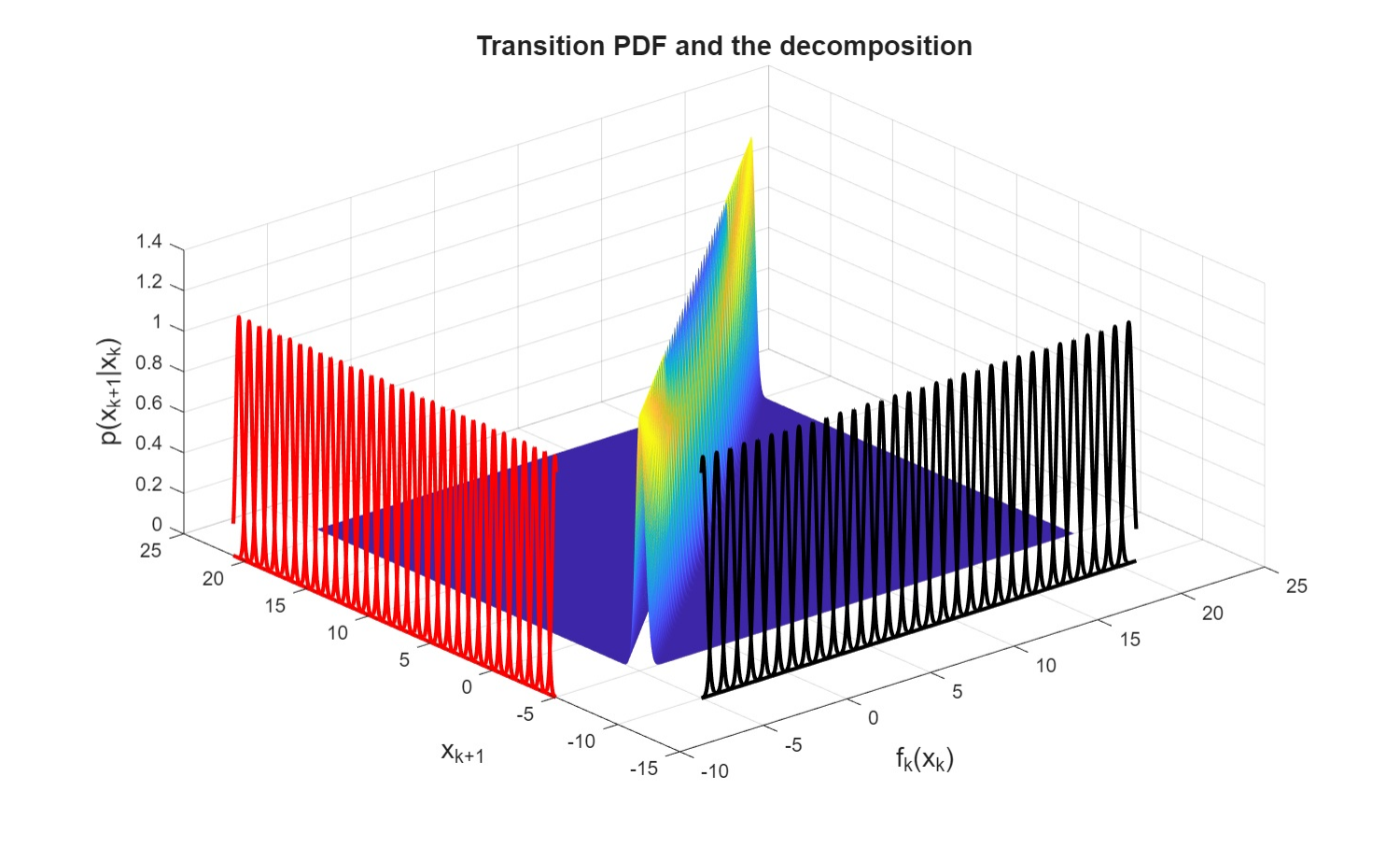}
    \caption{Illustration of the transition \ac{pdf} decomposition with \ac{psg}. Red curves correspond to functions $\bfg$ while black curves to functions $\bfphi$.}
    \label{fig:illustration_predicted}
\end{figure}

This decomposition assumes the transition density~\eqref{eq:transition_gaussian} decomposed as
\begin{align}\label{eq:decomposition_predicted}
    p(\bfx_{k+1}|\bfx_{k}) &\approx \sum_{j=1}^{M_{k+1}}\omega^j\, \bfg^j(\bfx_{k+1};\bftheta^{\bfg,j}_{k+1})\,\bfphi^j(\bff_{k}(\bfx_{k});\bftheta^{\bfphi,j}_k),
\end{align}
where functions $\bfg^j$ and $\bfphi^j$ are given by Gaussian \ac{pdf}
\begin{subequations}\label{eq:psg_decomposition}
\begin{align}\label{eq:psg_decomposition_xkpo}
    \bfg^j(\bfx_{k+1};\bftheta_{k+1}^{\bfg,j})&=\calN\{\bfx_{k+1};\bfm^{\bfg,j}_{k+1},\bfSigma^{\bfg,j}\}\\
    \label{eq:psg_decomposition_fxk}
    \bfphi^j(\bff_{k}(\bfx_{k});\bftheta_{k}^{\bfphi,j})&=\calN\{\bff_{k}(\bfx_{k});\bfm^{\bfphi,j}_k;\bfSigma^{\bfphi,j}\}
\end{align}
\end{subequations}
The decomposition~\eqref{eq:decomposition_predicted} was proposed in~\cite{TiStraDu:23} to address the solution to the \ac{cke} for the \ac{pmf}.
The decomposition defines the fixed structure of the predictive PDF $p(\bfx_{k+1}|\bfz^{k})$ through parameters 
$\bftheta^{\bfg}_{k+1}\coloneqq\{\bftheta^{\bfg,j}_{k+1}\}_{j=1}^{M_{k+1}}$, $\bftheta^{\bfphi}_{k}\coloneqq\{\bftheta^{\bfphi,j}_{k}\}_{j=1}^{M_{k+1}}$ and the weights $\omega_k\coloneqq\{\omega^j_k\}_{j=1}^{M_{k+1}}$.
The decomposition is illustrated in Figure~\ref{fig:illustration_predicted} for~\eqref{eq:ungm_transition}. Note that compared to the illustration in Figure~\ref{fig:illustration_filtered}, the axis-aligned grid is defined over the space of $x_{k+1}$ and $f_k(x_k)$ rather than $x_{k+1}$ and $x_k$, which is the case of \ac{fsg}.
\subsubsection{Calculation of the parameters}
The parameters are also obtained by an optimization process~\cite{TiStraDu:23}.
The number of terms $M_{k+1}$ and the locations $\bfm^{\bfg,j}_{k+1}$ are selected so that the approximation is good on the non-negligible support of the prior $p(\bfx_k|\bfz^k)$. The locations $\bfm^{\bfg,j}_{k+1}$ and $\bfm^{\bfphi,j}_{k}$ are usually equal $\bfm^{\bfg,j}_{k+1}=\bfm^{\bfphi,j}_{k}$, and variances $\bfSigma^{\bfg,j}$ and $\bfSigma^{\bfphi,j}$ are usually fixed and identical. They only depend on how dense the locations are.
For optimal parameter values, the grid can easily be redefined over a hyper-rectangular region of arbitrary size.
\subsubsection{Possible generalization}
The decomposition with \ac{psg}~\eqref{eq:decomposition_predicted} can be straightforwardly generalized to higher dimensions as demonstrated in~\cite{TiStraDu:23}. Similarly to the decomposition with the \ac{fsg}, the number of components grows exponentially with the dimension of the state $n_x$. The decomposition can be calculated for other classes of transition \ac{pdf} such as Student-t distributed \ac{pdf}~\cite{TiStraDu:23}.
Time-varying function $\bff_k$ in~\eqref{eq:sseqx} does not affect the decomposition since the decomposition is computed in the space of $\bfx_{k+1}$ and $\bff_k(\bfx_k)$ as illustrated in Figure~\ref{fig:illustration_predicted}. Also, time-varying parameters of the transition \ac{pdf} may not complicate the application of the decomposition since, for example, of Gaussian \ac{pdf}, the decomposition was calculated for standard Gaussian distribution and the parameters $\bftheta^{\bfg}_{k+1}$ and $\bftheta^{\bfphi}_{k}$ can easily be adapted for arbitrary covariance matrix.
\section{\acp{gmf} with transition PDF decompositions}\label{sec:GMM_TPD}
Now, the \ac{gmf} algorithms employing the decompositions~\eqref{eq:decomposition_filtered} and~\eqref{eq:decomposition_predicted} are introduced.
\subsection{\Ac{gmf} with transition PDF decomposition and \ac{fsg}}
The algorithm denoted as GMF-FSGD can be described using Algorithm~\ref{alg:gmf_fsg}.
\begin{algorithm}
\caption{GMF-FSGD}\label{alg:gmf_fsg}
\textbf{Step 1 (initialization):} Assume a posterior \ac{pdf}
\begin{subequations}
\begin{align}
p(\bfx_{k}|\bfz^{k})=\sum_{i=1}^{N_{k|k}}\alpha_{k|k}^i\,\calN\{\bfx_k;\bfm^{\bfx,i}_{k|k},\bfSigma^{\bfx,i}_{k|k}\}.
\end{align}
\textbf{Step 2 (time update):}  Given the support of the posterior, select the parameters $\bftheta^{\bfgamma}_{k}$, $\bftheta^{\bfg}_{k+1}$, and weight $\omega_k$.
Calculate the CKE as
\begin{multline}
    p(\bfx_{k+1}|\bfz^{k})=\int p(\bfx_{k+1}|\bfx_{k})p(\bfx_{k}|\bfz^{k})\d \bfx_k\\
    \approx \int \sum_{j=1}^{M_{k+1}}\omega^j_k\, \bfg^j(\bfx_{k+1};\bftheta^{\bfg,j}_{k+1})\,\bfgamma^j(\bfx_{k};\bftheta^{\bfgamma,j}_k)\\
    \sum_{i=1}^{N_{k|k}}\alpha_{k|k}^i\,\calN\{\bfx_k;\bfm^{\bfx,i}_{k|k},\bfSigma^{\bfx,i}_{k|k}\}\d\bfx_k,
\end{multline}
which can be rewritten as 
\begin{align}
p(\bfx_{k+1}|\bfz^{k})\approx\sum_{j=1}^{M_{k+1}}\beta^j\bfg^j(\bfx_{k+1};\bftheta^{\bfg,j}_{k+1}),
\end{align}
where
\begin{multline}\label{eq:gmf_fsg_beta}
    \beta^j=\omega^j_k\sum_{i=1}^{N}\alpha_{k|k}^i\int \bfgamma^j(\bfx_{k};\bftheta^{\bfgamma,j}_k)\,\calN\{\bfx_k;\bfm^{\bfx,i}_{k|k},\bfSigma^{\bfx,i}_{k|k}\}\d\bfx_k\\
    =\omega^j_k\sum_{i=1}^{N}\alpha_{k|k}^i\,\calN\{\bfm^{\bfgamma,j}_k;\bfm^{\bfx,i}_{k|k},\bfSigma^{\bfgamma,j}_k+\bfSigma^{\bfx,i}_{k|k}\}.
\end{multline}

Then, the predictive PDF $p(\bfx_{k+1}|\bfz^{k})$ is of the form
\begin{align}\label{eq:gmf_fsg_predictive}
p(\bfx_{k+1}|\bfz^{k})=\sum_{j=1}^{N_{k+1|k}}\alpha_{k+1|k}^j\,\calN\{\bfx_{k+1};\bfm^{\bfg,j}_{k+1},\bfSigma^{\bfg,j}\}
\end{align}
with $N_{k+1|k}=M_{k+1}$ and $\alpha_{k+1|k}^j=\beta^j$.
\end{subequations}

\textbf{Step 3 (measurement update):} Update each parameter of $p(\bfx_{k+1}|\bfz^{k})$ using standard equations for \ac{gmf}.
\end{algorithm}

Note that for the transition \ac{pdf} decomposition with \ac{fsg}, no approximation is used when calculating the \ac{cke} since the function $\bfgamma$ is a Gaussian \ac{pdf} in $\bfx_k$. The prediction \ac{pdf} in~\eqref{eq:gmf_fsg_predictive} demonstrates that its structure is defined by the user through the locations $\bfm_{k+1}^{\bfg,j}$.
\subsection{\Ac{gmf} with transition PDF decomposition and \ac{psg}}
The algorithm denoted as GMF-PSGD can be described using Algorithm~\ref{alg:gmf_psg}.
\begin{algorithm}
\caption{GMF-PSGD}\label{alg:gmf_psg}
\textbf{Step 1 (initialization):} Assume a posterior \ac{pdf}
\begin{subequations}
\begin{align}
    p(\bfx_{k}|\bfz^{k})=\sum_{i=1}^{N_{k|k}}\alpha_{k|k}^i\,\calN\{\bfx_k;\bfm^{\bfx,i}_{k|k},\bfSigma^{\bfx,i}_{k|k}\}.
\end{align}
\textbf{Step 2 (time update):}  Given the support of the posterior, select the parameters $\bftheta^{\bfphi}_k$, $\bftheta^{\bfg}_{k+1}$, and the weights $\omega_k$.
Calculate the CKE as
\begin{multline}
    p(\bfx_{k+1}|\bfz^{k})=\int p(\bfx_{k+1}|\bfx_{k})p(\bfx_{k}|\bfz^{k})\d \bfx_k\\
    \approx \int \sum_{j=1}^{M_{k+1}}\omega^j_k\, \bfg^j(\bfx_{k+1};\bftheta^j_{k+1})\,\bfphi^j(\bff_{k}(\bfx_{k});\bftheta^{\bfphi,j}_k)\\
    \sum_{i=1}^{N_{k|k}}\alpha_{k|k}^i\,\calN\{\bfx_k;\bfm^{\bfx,i}_{k|k},\bfSigma^{\bfx,i}_{k|k}\}\d\bfx_k,\\
\end{multline}
which can be rewritten as 
\begin{align}
    p(\bfx_{k+1}|\bfz^{k})\approx\sum_{j=1}^{M_{k+1}}\beta^j\,\bfg^j(\bfx_{k+1};\bftheta^{\bfg,j}_{k+1}),
\end{align}
where
\begin{multline}\label{eq:gmf_psg_beta}
    \beta^j=\omega^j\sum_{i=1}^{N_{k|k}}\alpha_{k|k}^i\\
    \int \bfphi^j(\bff_{k}(\bfx_{k});\bftheta^{\bfphi,j}_k)\,\calN\{\bfx_k;\bfm^{\bfx,i}_{k|k},\bfSigma^{\bfx,i}_{k|k}\}\d\bfx_k.
\end{multline}
Then, the predictive PDF $p(\bfx_{k+1}|\bfz^{k})$ is of the form
\begin{align}
    p(\bfx_{k+1}|\bfz^{k})=\sum_{j=1}^{N_{k+1|k}}\beta^j\,\calN\{\bfx_{k+1};\bfm^{\bfg,j}_{k+1},\bfSigma^{\bfg,j}\}
\end{align}
with $N_{k+1|k}=M_{k+1}$.
\end{subequations}

\textbf{Step 3 (measurement update):} Update each parameter of $p(\bfx_{k+1}|\bfz^{k})$ using standard equations for \ac{gmf}.
\end{algorithm}
For generally nonlinear $\bff_k$, the integral in \eqref{eq:gmf_psg_beta} has to be evaluated approximately.
The approximation error should be small since both variances $\bfSigma^{\bfphi,j}_k$ and $\bfSigma^{\bfx,i}_{k|k}\preceq \bfSigma^{\bfg,j}_k$ are small and the effect of the nonlinearity is thus limited. They are also identical by design, so the calculations can be simple.
\subsection{Theoretical comparison of GMF-FSGD and GMF-PSGD}
Both algorithms are based on a decomposition of the transition \ac{pdf}, and their application for a state estimation problem must involve an offline stage, where the decomposition is pre-calculated.
The comparison thus concentrates on both aspects: the offline precomputation stage and the algorithm application for online state estimation. Both grids are defined as axis-aligned. While \ac{fsg} is defined over the space given by $\bfx_{k+1}$ and $\bfx_k$, the \ac{psg} is defined over the space defined by  $\bfx_{k+1}$ and $\bff_k(\bfx_k)$.

For the \ac{fsg}, the offline stage involves the specification of a grid in the domain of $\bfx_k$. The grid has to cover the region where the state $\bfx_k$ is expected to lie for $k=1,2,\ldots$. Defining the region may be challenging, and for real-world problems, one needs to run a simple filter first to provide a rough state estimate to determine the region. The size of the region substantially affects the number of terms of the decomposition. 
The online state estimation does not necessarily use all the terms of the decomposition (precomputed for all time instants) but rather a subset of terms relevant to the current state estimate $\hbfx_{k|k}$. As mentioned above, the time-dependency of the function $\bff_k(\bfx_k)$ in~\eqref{eq:sseqx} may complicate the decomposition precalculation further. 

All the difficulties of \ac{fsg} stem from the definition of the grid over the space given by $\bfx_{k+1}$ and $\bfx_{k}$. Since the \ac{psg} is defined over the space defined by $\bfx_{k+1}$ and $\bff_k(\bfx_k)$, it is only the process noise distribution $p_{\bfw_k}$ that affects the \ac{psg}. Moreover, even if the distribution parameters vary in time, this dependency may not complicate the \ac{psg} computation. An example is the Gaussian $\bfw_k$ with a time-varying mean or covariance matrix, where the \ac{psg} can be precomputed for a standard Gaussian distribution and then shifted and scaled to align with $p_{\bfw_k}$. 

\begin{figure*}
    \centering
    \includegraphics[width=\linewidth]{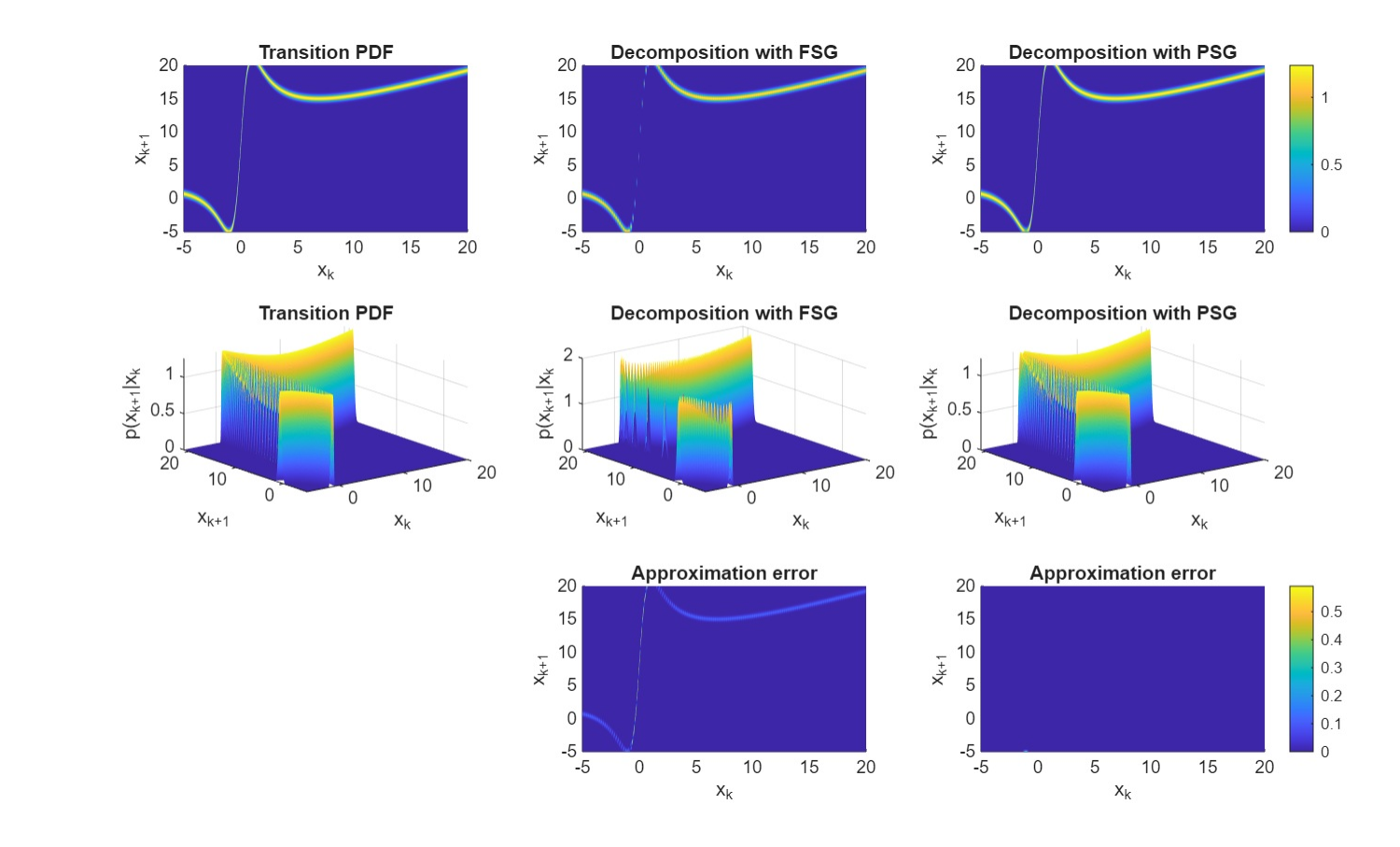}
    \caption{Comparison of transition \ac{pdf} and its approximations for the \ac{ungm}: top row -- top view, middle row -- 3D view, bottom row -- square of approximation error.}
    \label{fig:comparison}
\end{figure*}

From the offline stage calculation point of view, the \ac{fsg} is more difficult to calculate than the \ac{psg}. Both grids are usually specified to be equidistant, which for the \ac{fsg} may lead to limited approximation accuracy in areas when small changes to $\bfx_k$ correspond to big changes in $\bfx_{k+1}$. This is illustrated in Figure~\ref{fig:comparison}, where the \ac{ungm} transition \ac{pdf} is decomposed for both \ac{fsg} and \ac{psg}.

The disadvantage of the decomposition with \ac{fsg} is compensated in the online state estimation, where due to the function $\bfgamma^j$~\eqref{eq:fsg_decomposition_xk} being Gaussian \ac{pdf} in $\bfx_k$, the integration in~\eqref{eq:gmf_fsg_beta} can be evaluated analytically. The GMF-PSGD has to calculate the corresponding integral \eqref{eq:gmf_psg_beta} numerically as the function $\bfphi^j$ in ~\eqref{eq:psg_decomposition_fxk} is a Gaussian \ac{pdf} in $\bff_k(\bfx_k)$ rather in $\bfx_k$. The numerical approximation used for the evaluation may be a simple one, e.g., the cubature rule, which is computationally cheap and leads to satisfactory results since both \acp{pdf}, which product appears in the integral, have similar covariance matrices.

\section{Numerical illustration}\label{sec:numerical_illustration}
Consider the \ac{ungm}~\cite{GoSaSmi:93}, which is strongly nonlinear and often used as a benchmark problem. Its transition \ac{pdf} is given in~\eqref{eq:ungm_transition} and measurement \ac{pdf}
\begin{align}
    p(\bfz_k|\bfx_k)=\calN\{\bfz_k;\tfrac{x_k^2}{20},R\}.
\end{align}
In this paper, we consider the process and measurement noise variances $Q=10^{-1}$ and $R=10^{-1}$, respectively, and the initial condition being Gaussian with zero mean and variance $\bfSigma^{\bfx}_{0|0}=10^{-2}$.

\begin{figure}[b]
    \centering
    \includegraphics[width=\linewidth,height=5cm]{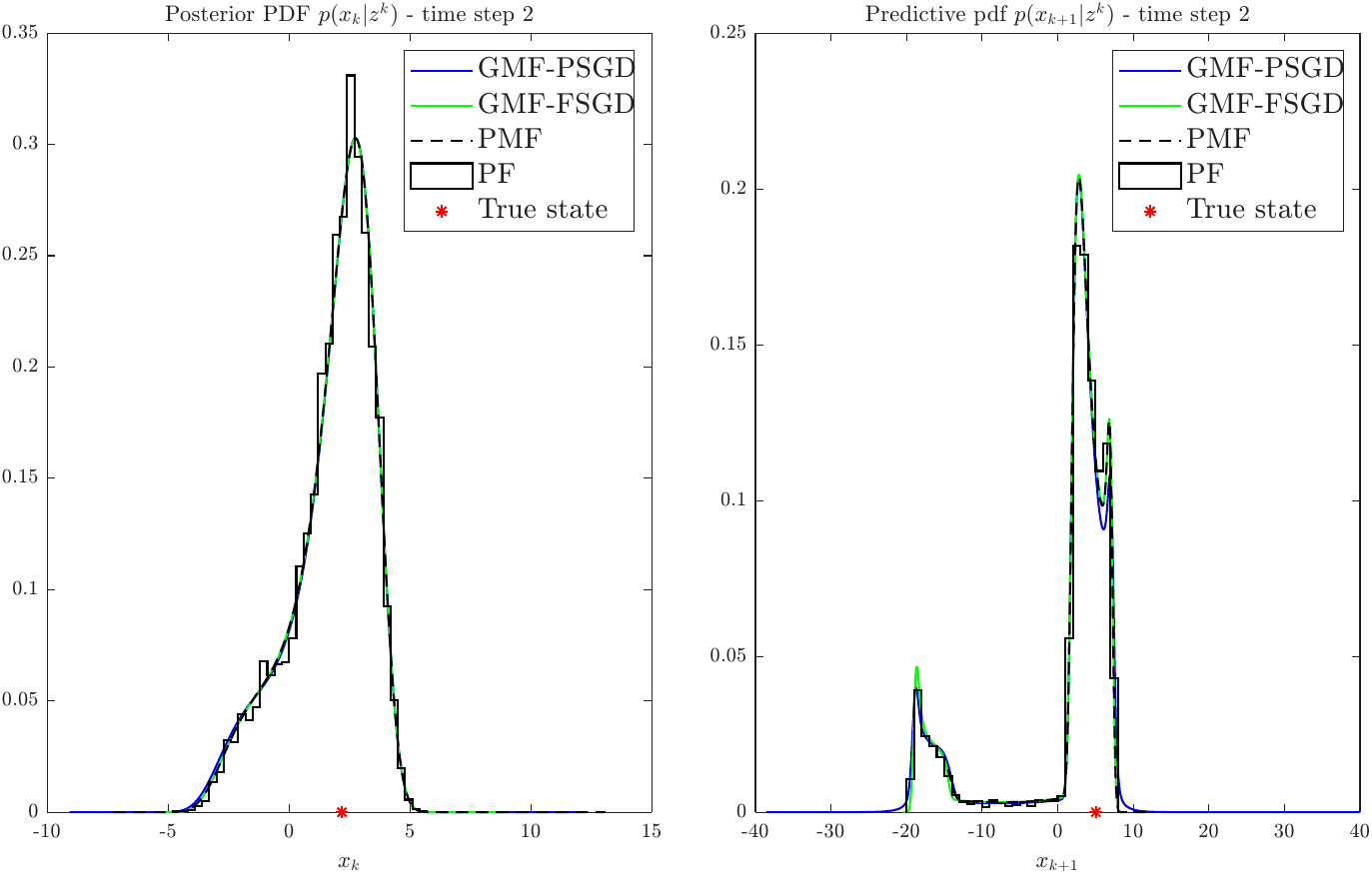}
    \caption{Illustration of the filtering and predictive \acp{pdf} produced by the filters ($k=2$).}
    \label{fig:step_ii}
\end{figure}

\begin{figure}
    \centering
    \includegraphics[width=\linewidth,height=4.5cm]{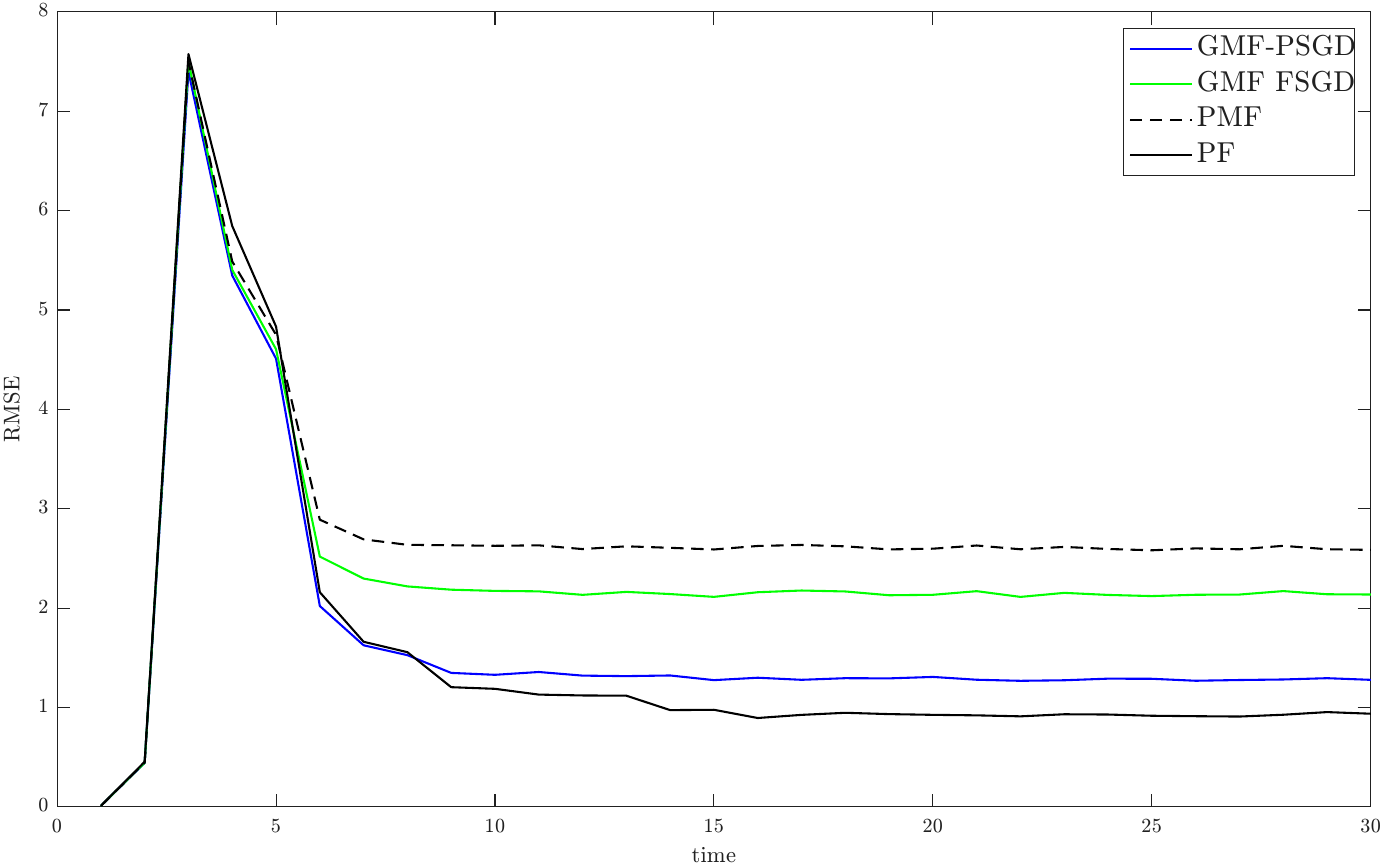}
    \caption{Time evolution of \ac{rmse}.}
    \label{fig:rmse}
\end{figure}

The state of the \ac{ungm} was estimated by
\begin{itemize}
    \item GMF-FSGD with equidistant locations   $m^{\bfgamma,j}$ with distance between neighborhood locations $0.05\sqrt{Q}$,
    \item GMF-PSGD with the decomposition parameters computed in~\cite{TiStraDu:23} with rank 40,
    \item \ac{pmf}~\cite{SiKraTi:01} with $10^3$ grid points,
    \item \ac{pf} with the importance density given by the transition \ac{pdf}, multinomial resampling at each time instant, and $N^{\pf}=10^3$ samples.
\end{itemize}
First, the predictive and filtering \acp{pdf} for $k=2$  are shown in Figure~\ref{fig:step_ii} 
to illustrate the capability of the proposed filters to produce a good approximation of complex-shaped \acp{pdf}.
Then, the \ac{rmse} 
\begin{align}
    \rmse_k = \sqrt{\sum_{\ell=1}^{M}\left(x_k(\ell)-\hx_{k|k}(\ell)\right)^2}
\end{align}
based on $M=10^4$ \ac{mc} simulations was computed with $x_k(\ell)$ being the true state at $\ell$-th \ac{mc} simulation and $\hx_{k|k}(\ell)$ being its filtering estimate. It is shown in Figure~\ref{fig:rmse}, from which it follows that for the above-mentioned parameters, the best \ac{rmse} performance has the \ac{pf}.
The GMF-PSGD performs better than the GMF-FSGD, which is probably given by the fact that the decomposition with the \ac{fsg} has worse approximation quality in some regions compared to the decomposition with the \ac{psg}.

\begin{figure}
    \centering
    \includegraphics[width=\linewidth,height=4.5cm]{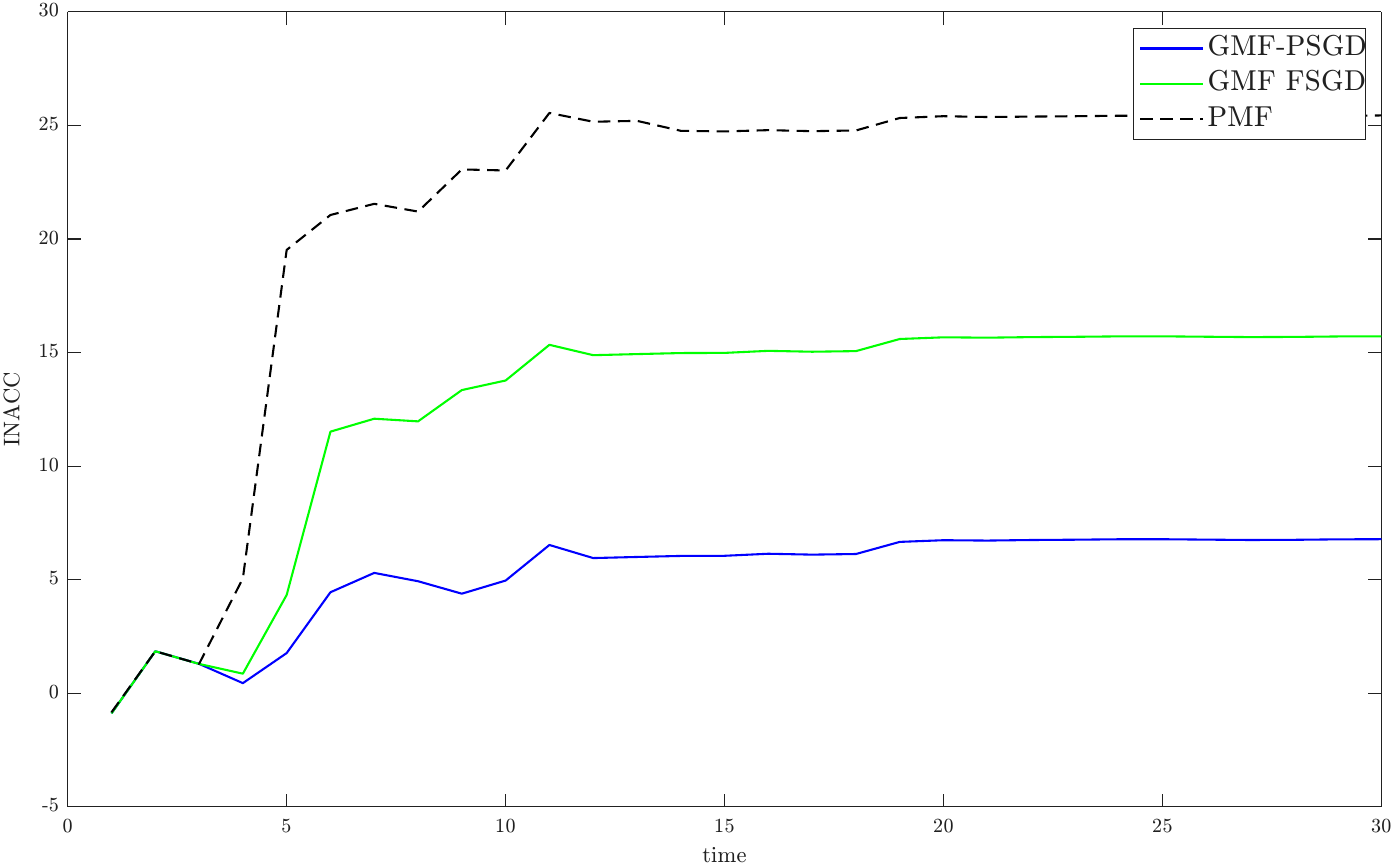}
    \caption{Time evolution of INACC.}
    \label{fig:inacc}
\end{figure}

Since the \ac{rmse} only evaluates the quality of point estimates, the proposed methods are also assessed based on the quality of the posterior \ac{pdf}.
The performance was analyzed in terms of \ac{kld} between the posterior \ac{pdf} produced by the  \ac{pf} with $N^\pf=10^5$ samples representing the true posterior and the posterior \acp{pdf} produced by GMF-PSGD, GMF-FSGD, and \ac{pmf}.
The posterior \ac{pdf} of the \ac{pf} is given as
\begin{align}
    p^\pf(\bfx_k|\bfz^k)=\tfrac{1}{N^\pf}\sum_{i=1}^{N^\pf}\delta(\bfx_k-\bfs_k^i),
\end{align}
where $\delta$ is the Dirac-delta function and $\bfs_k^i$ is the $i$-th sample.
The \ac{kld} defined as
\begin{align}
  D_{\mathrm{KL}}(p^\pf\|p)&=\int p^\pf(\bfx_k|\bfz^k)\log\frac{p^\pf(\bfx_k|\bfz^k)}{p(\bfx_k|\bfz^k)}\d\bfx_k\nonumber\\
  &=\underbrace{\int p^\pf(\bfx_k|\bfz^k)\log p^\pf(\bfx_k|\bfz^k)\d\bfx_k}_{\text{SDE}}\nonumber\\
  &\underbrace{- \int p^\pf(\bfx_k|\bfz^k)\log p(\bfx_k|\bfz^k)\d\bfx_k}_{\textbf{INACC}} 
\end{align}
consists of \ac{sde} and inaccuracy (INACC)~\cite{Ku:96}.
The discrepancy between the \acp{pdf} is captured by the inaccuracy only; hence, it will be used for the comparison.
The inaccuracy averaged over the \ac{mc} simulations is depicted in Figure~\ref{fig:inacc}.
From the figure, it follows that superior performance is achieved by the GMF-PSGD while the \ac{pmf} does not perform well.

The computational time of the algorithms is compared in terms of the time of the filter's single step, given in Table~\ref{tab:time}

\begin{table}[ht]
    \centering
    \begin{tabular}{ccccc}
    \toprule
    & GMF-PSGD&GMF-FSGD&PF& PMF\\
    \midrule    
     Time [ms]& 0.78 & 1.96 & 4.29 & 6.09\\
    \bottomrule
    \end{tabular}
    \caption{Computational costs of the algorithms (single step of online part)}
    \label{tab:time}
\end{table}
The results indicate that the GMF-PSGD algorithm is computationally cheaper than GMF-FSGD. This was not expected since GMF-PSGD uses numerical integration while GMF-FSGD uses analytical integration in the online run. The reason for this is the number of terms, which was higher for GMF-FSGD (286 GM terms on average) than for GMF-PSGD (80 GM terms on average). The difference is caused by different approximations and sizes of the approximation regions used by the algorithms.
It shall be noted that all the above results depend on the specification of the filter parameters, such as the number of grid points, samples, and \ac{gm} terms in the transition \ac{pdf} decomposition. 
Comparing the performance of the filters is challenging due to differing parametrization.
Therefore, the purpose of the numerical example is to showcase the competitiveness of the proposed algorithms when compared with the representative Bayesian filtering algorithms.
\section{Conclusion}\label{sec:conclusion}

The paper proposed practical Gaussian mixture filters that do not rely on local component processing and do not suffer from component explosion. The key idea of these \emph{global} filters is the offline decomposition of a given transition density into a mixture of axis-aligned Gaussian components. These decompositions automatically maintain a predefined number of posterior components without requiring explicit component reduction. Two types of decompositions are derived that differ in the definition of the coordinate axes. While the first decomposition uses $\bfx_{k+1}$ and $\bfx_k$ as axes, the second one uses $\bfx_{k+1}$ and $\bff(\bfx_k)$. The first decomposition has complex offline processing and requires the specification of an approximation domain. However, for Gaussian and Gaussian mixture posteriors, the prediction step can then be performed online in closed form.
The second decomposition is simple to perform and does not require a predefined approximation domain but requires numerical integration (done by the unscented transform). 
Both decompositions allow an adjustable trade-off between the computational load during the online prediction step and the estimation quality achieved. Numerical simulations with a highly nonlinear univariate nonstationary Gaussian model (UNGM) problem demonstrated that the proposed algorithms are competitive compared to the state-of-the-art Bayesian filtering methods in terms of both point estimate quality and posterior \ac{pdf} quality.

\textbf{Future work} will focus on the calculation of more efficient decompositions for GMF-FSGD and its generalization to higher dimensions, as GMF-PSGD can easily be extended to mid dimensions (approx. ten state elements).
\bibliographystyle{IEEEtran}

\end{document}